\documentclass[prl,aps,floats,twocolumn]{revtex4}

\usepackage[dvips]{graphicx}
\usepackage{amssymb} 
\usepackage{amsmath}
\usepackage{ctable}
\usepackage{url}
\usepackage{color}

\begin{document}

\title{The Standard Model, The Exceptional Jordan Algebra, and Triality}

\author{Latham Boyle} 

\affiliation{ 
 Higgs Centre for Theoretical Physics, University of Edinburgh, EH9 3FD, UK \\
 Perimeter Institute for Theoretical Physics, Waterloo, Ontario, N2L 2Y5, Canada}

\date{June 2020}
\begin{abstract}
Jordan, Wigner and von Neumann classified the possible algebras of quantum mechanical observables, and found they fell into 4 ``ordinary" families, plus one remarkable outlier: the exceptional Jordan algebra.  We point out an intriguing relationship between the complexification of this algebra and the standard model of particle physics, its minimal left-right-symmetric $SU(3)\times SU(2)_{L}\times SU(2)_{R}\times U(1)$ extension, and $Spin(10)$ unification.  This suggests a geometric interpretation, where a single generation of standard model fermions is described by the tangent space $(\mathbb{C}\otimes\mathbb{O})^{2}$ of the complex octonionic projective plane, and the existence of three generations is related to $SO(8)$ triality.
\end{abstract}

\maketitle

\section{Introduction}  

Many basic questions about the standard model of particle physics remain open.  For example:
\begin{itemize}
\item (i) Where does the standard model gauge group $G_{SM}=[SU(3)_{C}\times SU(2)_{L}\times U(1)_{Y}]/\mathbb{Z}_{6}$ come from?  
\item (ii) Why does each generation of standard model fermions (including a right-handed neutrino in each generation) transform according to the following strange representation of $G_{SM}$:
\begin{eqnarray}
  \label{rho_SM}
  \rho_{SM}\!&\!=\!&\!(3,2,\!+\frac{1}{6})\oplus(\bar{3},1,\!+\frac{1}{3})\oplus(\bar{3},1,\!-\frac{2}{3}) \nonumber \\
  \!&\!\oplus\!&\!(1,2,\!-\frac{1}{2})\oplus(1,1,\!+\;\!1\;\!)\oplus(\;\!1,1,0)\;?
\end{eqnarray}
\item (iii) Why are there three generations of standard model fermions ({\it i.e.}\ three copies of $\rho_{SM}$)?
\end{itemize}

Soon after the standard model fell into place, physicists had an important insight into question (ii): they noticed that if $G_{SM}$ were embedded in certain simpler, more unified groups like $SU(5)$ \cite{Georgi:1974sy} or $Spin(10)$ \cite{Georgi:1974my, Fritzsch:1974nn}, then the standard model fermions would simultaneously organize themselves into simpler, more unified representations of those groups (see \cite{Baez:2009dj} for an introduction).  Most strikingly, they found that if $G_{SM}$ were embedded in the simple group $Spin(10)$, then each generation of 16 standard model fermions would also be precisely described by a {\it single} 16-dimensional complex irrep (the Weyl spinor) of $Spin(10)$ so that, after restricting $Spin(10)$ to its subgroup $G_{SM}$, this single irrep would decompose precisely into the desired representation $\rho_{SM}$!  It is hard to believe that this is a coincidence, and more likely that it is an important clue about the standard model and what lies beyond it, although the correct interpretation of this clue (and, in particular, whether it points to a grand-unified gauge theory in four-dimensional spacetime, as physicists originally believed in the 1970s) is still unclear. 

Recently, as we review in the next section, Todorov and Dubois-Violette \cite{Todorov:2018mwd} pointed out an intriguing new connection between the standard model and the exceptional Jordan algebra $h_{3}(\mathbb{O})$, which may be a new clue about questions (i) and (iii).  Indeed, they propose $h_{3}(\mathbb{O})$ as the structure underlying the standard model \cite{Dubois-Violette:2016kzx, Todorov:2018yvi, Dubois-Violette:2018wgs, Todorov:2019hlc, Dubois-Violette:2020hpk}.

Here we make a new proposal, inspired by their clue.  In brief, we suggest that $h_{3}(\mathbb{O})$ should be replaced by its {\it complexification} $h_{3}^{C}(\mathbb{O})$.  As we will see, this change: (i) incorporates the earlier $Spin(10)$ insight; (ii) embeds $G_{SM}$ and $\rho_{SM}$ in a left-right (``LR") symmetric extension based on $SU(3)\!\times\!SU(2)_{L}\!\times\!SU(2)_{R}\!\times\!U(1)$ which (as we review) has recently been shown to neatly explain several observational facts \cite{Hall:2018let, Hall:2019qwx, Dror:2020jzy}); and (iii) suggests a new geometric interpretation of the standard model fermions, based on $E_{6}$, triality, and the ``magic square."

\section{Review: $G_{SM}$ from $h_{3}(\mathbb{O})$.}  

In this section, we review the main result in \cite{Todorov:2018mwd}, and its interpretation \cite{BaezBlogPost}.  

First, we must introduce the octonions $\mathbb{O}$ and the exceptional Jordan algebra $h_{3}(\mathbb{O})$.  For a more in-depth introduction to these topics, see the excellent review \cite{Baez:2001dm}.

\subsection{The octonions $\mathbb{O}$.}  

\begin{figure}
  \begin{center}
    \includegraphics[width=1.9in]{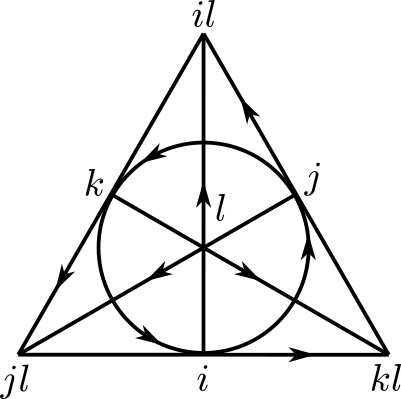}
  \end{center}
  \caption{Fano plane summarizing octonionic multiplication, and our conventions.  See Sec. 2.1 in \cite{Baez:2001dm} for more explanation.}
 \label{FanoPlane}
\end{figure}

A celebrated theorem states that there are only four normed division algebras: the real numbers $\mathbb{R}$, the complex numbers $\mathbb{C}$, the quaterions $\mathbb{H}$, and the octonions $\mathbb{O}$ (which are 1-, 2-, 4-, and 8-dimensional, respectively).  Just as a complex number $z\in\mathbb{C}$ may be written $z=a_{0}+a_{1}i$, and a quaternion $q\in\mathbb{H}$ may be written $q=a_{0}+a_{1}i+a_{2}j+a_{3}k$, an octonion $x\in\mathbb{O}$ may be written 
\begin{equation}
  \label{octonion}
  x=a_{0}+a_{1}i+a_{2}j+a_{3}k+a_{4}l+a_{5}\;\!il+a_{6}\;\!jl+a_{7}\;\!kl
\end{equation}
where the $a_{\mu}$ ($\mu=0,\ldots,7$) are real coefficients, each of the seven imaginary units $\{i,j,k,l,il,jl,kl\}$ squares to $-1$, and the product of any two {\it distinct} imaginary units gives a third imaginary unit, according to the rule shown by the ``Fano plane" in Fig.~\ref{FanoPlane}.  The octonion $x$ has conjugate $x^{\ast}$ given by negating the $a_{i}$ (for $i=1,\ldots,7$).

\subsection{The exceptional Jordan algebra $h_{3}(\mathbb{O})$.}  

Soon after the discovery of quantum mechanics, Jordan characterized the possible algebras of quantum mechanical observables; and these ``Euclidean Jordan algebras" were then classified by Jordan, Wigner and von Neumann (JWvN) \cite{Jordan:1933vh}.  Much as Killing and Cartan classified the simple Lie algebras into 4 infinite families ($A_{n}$, $B_{n}$, $C_{n}$ and $D_{n}$) and 5 exceptional cases ($G_{2}$, $F_{4}$, $E_{6}$, $E_{7}$ and $E_{8}$), JWvN classified the simple Euclidean Jordan algebras into 4 infinite families and {\it one} outlier: the exceptional Jordan algebra.  This is the algebra $h_{3}(\mathbb{O})$ of hermitian $3\times3$ matrices with octonionic entries
\begin{equation}
  \label{Jordan_element}
  y=\left(\begin{array}{ccc}
  \alpha_{1} & x_{3} & x_{2}^{\ast} \\
  x_{3}^{\ast} & \alpha_{2} & x_{1} \\
  x_{2} & x_{1}^{\ast} & \alpha_{3} \end{array}\right)\qquad(\alpha_{i}\in\mathbb{R},\;\;x_{i}\in\mathbb{O})
\end{equation}
and with the product $y_{1}\circ y_{2}$ of two elements $y_{1},y_{2}\in h_{3}(\mathbb{O})$ given by half their matrix anti-commutator:
\begin{equation}
  y_{1}\circ y_{2}\equiv\frac{1}{2}(y_{1}y_{2}+y_{2}y_{1}).
\end{equation}

\subsection{Automorphisms of $h_{3}(\mathbb{O})$.}  

The automorphisms of $h_{3}(\mathbb{O})$ form the exceptional Lie group $F_{4}$ \cite{Baez:2001dm, Yokota}.  Among these automorphisms, we now describe two natural subgroups, $H_{1}$ and $H_{2}$, which intersect to form $G_{SM}$.

To describe the first subgroup, $H_{1}$, we start by choosing an embedding of $\mathbb{C}$ in $\mathbb{O}$: without loss of generality, we choose the imaginary unit $l\in\mathbb{O}$ to coincide with the imaginary unit in $\mathbb{C}$.  Then, writing the octonion (\ref{octonion}) as
\begin{equation}
  x=(a_{0}+a_{4}l)+i(a_{1}+a_{5}l)+j(a_{2}+a_{6}l)+k(a_{3}+a_{7}l)
\end{equation}
we can associate each element $x\in\mathbb{O}$ with an element $z+\vec{Z}\in\mathbb{C}\oplus\mathbb{C}^{3}$, where
\begin{equation}
  z\equiv a_{0}+a_{4}l\in\mathbb{C},\qquad
  \vec{Z}\equiv \left(\begin{array}{c}
  a_{1}+a_{5}l \\ a_{2}+a_{6}l \\ a_{3}+a_{7}l \end{array}\right)\in\mathbb{C}^{3}.
\end{equation}
This same choice associates each of the three octonions $x_{i}\in\mathbb{O}$ in (\ref{Jordan_element}) with an element $z_{i}+\vec{Z}_{i}\in\mathbb{C}\oplus\mathbb{C}^{3}$, so that we can associate the 
element $y\in h_{3}(\mathbb{O})$ with an element $\hat{y}+m\in h_{3}(\mathbb{C})\oplus M_{3}(\mathbb{C})$ where
\begin{equation}
  \hat{y}\equiv\!\left(\!\!\begin{array}{ccc}
  \alpha_{1} & \!z_{3}\! & z_{2}^{\ast} \\
  z_{3}^{\ast} & \!\alpha_{2}\! & z_{1} \\
  z_{2} & \!z_{1}^{\ast}\! & \alpha_{3} \end{array}\!\!\right)\!\!\in\!h_{3}(\mathbb{C}),\;\;\,
  m\!\equiv\!(\vec{Z}_{1}\vec{Z}_{2}\vec{Z}_{3})\!\in\!M_{3}(\mathbb{C}).
\end{equation}
The general automorphism of $h_{3}(\mathbb{O})$ that preserves this embedding of $\mathbb{C}$ in $\mathbb{O}$ is then (see Thrm. 2.12.2 in  \cite{Yokota})
\begin{equation}
  \label{H1}
  \hat{y}\to V\hat{y}V^{\dagger},\quad m\to UmV^{\dagger}
\end{equation}
where $U$ and $V$ are arbitrary elements of $SU(3)$.  Since the automorphism is unchanged by the transformation $\{U,V\}\to {\rm e}^{2\pi i/3}\{U,V\}$, these automorphisms form the subgroup $H_{1}=[SU(3)\times SU(3)]/\mathbb{Z}_{3}$.

To describe the second subgroup, $H_{2}$, note that there is a duality between the exceptional Jordan algebra $h_{3}(\mathbb{O})$ and a geometric object: the octonionic projective plane (or ``Moufang plane") 
$\mathbb{O}\mathbb{P}^{2}$  \cite{Moufang, Baez:2001dm}.  In particular, the points in $\mathbb{O}\mathbb{P}^{2}$ correspond to the rank-one idempotents in $h_{3}(\mathbb{O})$, and the isometries of $\mathbb{O}\mathbb{P}^{2}$ correspond to the automorphisms of $h_{3}(\mathbb{O})$.  In particular, the isometries of $\mathbb{O}\mathbb{P}^{2}$ that fix a point correspond to the automorphisms of $h_{3}(\mathbb{O})$ which fix a rank-one idempotent $\Pi$ which, without loss of generality, we can take to be
\begin{equation}
  \label{p1}
  \Pi=\left(\begin{array}{ccc} 0 & 0 & 0 \\ 0 & 0 & 0 \\ 0 & 0 & 1 \end{array}\right).
\end{equation}
These form the subgroup $H_{2}\!=\!Spin(9)$ (Thrm 2.7.4 \cite{Yokota}).

\subsection{Todorov and Dubois-Violette's result.}  The intersection of these two subgroups consists of the automorphisms that simultaneously preserve the embedding of $\mathbb{C}$ in $\mathbb{O}$, and also 
preserve the idempotent $\Pi$.  These are the automorphisms of the form (\ref{H1}) with the additional property that $V\in SU(3)$ has the block diagonal form
\begin{equation}
  V=\left(\begin{array}{c|c} \varphi v & 0 \\ \hline 0 & \varphi^{-2} \end{array}\right)
\end{equation}
where $v\in SU(2)$ and $\varphi\in U(1)$.  Thus, an element $\{U,v,\varphi\}\in SU(3)\times SU(2)\times U(1)$ determines an automorphism in $H_{1}\cap H_{2}$; but this automorphism is unchanged by the transformation $\{U,v,\varphi\}\to\{U,-v,-\varphi\}$ or by the transformation $\{U,v,\varphi\}\to\{{\rm e}^{2\pi i/3}U,v,{\rm e}^{2\pi i/3}\varphi\}$.  So the intersection $H_{1}\cap H_{2}$ is precisely the standard model gauge group $G_{SM}=[SU(3)\times SU(2)\times U(1)]/\mathbb{Z}_{6}$, which is the key result in Ref.~\cite{Todorov:2018mwd}.
 
\subsection{Baez's Interpretation.} Baez gave a suggestive interpretation of this result \cite{BaezBlogPost}.  First note that the automorphisms in $H_{2}$ are those that fix a copy of $h_{2}(\mathbb{O})$ within $h_{3}(\mathbb{O})$, while those in $H_{1}\cap H_{2}=G_{SM}$ {\it also} fix a copy of $h_{2}(\mathbb{C})$ within $h_{2}(\mathbb{O})$.  Next note that $h_{2}(\mathbb{O})$ is a 10-dimensional vector space, and the $2\times2$ matrix determinant equips this space with a natural quadratic form of signature $(1,9)$.  In other words, $h_{2}(\mathbb{O})$ may be identified with 10d Minkowski spacetime $M_{10}$;
and, by the same token, $h_{2}(\mathbb{C})$ may be identified with 4d Minkowski spacetime $M_{4}$.  In this sense, we may interpret Todorov and Dubois-Violette's result \cite{Todorov:2018mwd} as saying: 

{\it If we fix a copy of $M_{10}$ inside $h_{3}(\mathbb{O})$, and also fix a copy of $M_{4}$ inside $M_{10}$, the residual symmetry is $G_{SM}$!}

This way of phrasing the result is strikingly suggestive.  Moreover, as reviewed in \cite{Baez:2001dm}, $SO(8)$ triality plays a fundamental underlying role in the exceptional Jordan algebra and, as emphasized in  \cite{Dubois-Violette:2016kzx, Todorov:2018yvi, Dubois-Violette:2018wgs, Todorov:2019hlc, Dubois-Violette:2020hpk}, it is tempting to speculate that this phenomenon underlies the existence of three generations of standard model fermions.  (We will have more to say about this below.)   

This completes our review of previous observations: taken together, they are an intriguing hint that the exceptional Jordan algebra is related to the standard model.   

\section{Left-right (LR) symmetry from $h_{3}^{C}(\mathbb{O})$.}  

Let us take this hint seriously, and see where it leads.  A natural question is how the standard model fermions, with their peculiar representation (\ref{rho_SM}), fit into the picture?  Refs.~\cite{Dubois-Violette:2016kzx, Todorov:2018yvi, Dubois-Violette:2018wgs, Todorov:2019hlc, Dubois-Violette:2020hpk} have made a few different proposals in this regard. (Also see \cite{Gunaydin:1973rs, Gunaydin:1974fb, Dixon, Furey:2018yyy, Boyle:2019cvm, Farnsworth:2020ozj} for various other proposals about how the octonions and/or Jordan algebras are related to the standard model.)  Here we are led to a different proposal.

To motivate our solution, let us start with a problem: we have seen that the symmetry underlying $h_{3}(\mathbb{O})$ (or its subalgebra $h_{2}(\mathbb{O})$) is $F_{4}$ (or its subgroup $Spin(9)$), and both of these groups contain $G_{SM}$.  But there is a reason that one ordinarily does not encounter unified theories based on these gauge groups: they only have real or pseudo-real representations, while the standard model representation (\ref{rho_SM}) is complex, and complex representations are needed in order to construct a chiral gauge theory.  (A ``complex" representation is one that is not equivalent to its complex conjugate representation.)  Extending from the automorphism groups ($F_{4}$ and $Spin(9)$) to the corresponding ``structure groups" ($E_{6,-26}$ and $Spin(9,1)$) is no help, as the relevant representations (${\bf 27}$ and ${\bf 16}$) are still not complex \cite{Baez:2001dm}. 

How, then, should we obtain the desired complex representation from the exceptional Jordan algebra?  A clue is that, among the five compact exceptional groups, $E_{6}$ is the only one with complex representations. (For this reason, $E_{6}$ has long been considered a candidate grand unified gauge group \cite{Gursey:1975ki}; but in our paper, $E_{6}$ appears in a different way.)  Can we obtain compact $E_{6}$ from the exceptional Jordan algebra?  Yes, but we must switch from $h_{3}(\mathbb{O})$ (a 27-dimensional algebra over the field $R$ of real scalars), to its {\it complexification} $h_{3}^{C}(\mathbb{O})$ (the corresponding 27-dimensional algebra over the field $C$ of complex scalars).  Indeed, $E_{6}$ is most elegantly defined as the group of invertible linear transformations from $h_{3}^{C}(\mathbb{O})$ to itself, that preserve its natural inner product and determinant (see Section 3.1 in \cite{Yokota}).  
 
Next we define subgroups ($\tilde{H}_{1}$ and $\tilde{H}_{2}$) of $E_{6}$ that parallel the previous section's subgroups ($H_{1}$ and $H_{2}$) of $F_{4}$.  

Before we describe these subgroups in detail, we must be careful to distinguish two different copies of the complex numbers: one copy $\mathbb{C}$ is embedded in $\mathbb{O}$ (with imaginary unit given as before by the imaginary octonion $l$), while the other copy $C$ is the field of complex scalars (whose imaginary unit we will denote by $I$).  We should also be careful to distinguish between three types of conjugation: conjugation in the field of complex scalars $C$, which we denote by $z\to\bar{z}$; conjugation in $\mathbb{O}$ (and hence also conjugation in $\mathbb{C}$) which we denote, as before, by $x\to x^{\ast}$; and conjugation in $\mathbb{O}$ combined with matrix transposition, which we denote by $A\to A^{\dagger}=(A^{\ast})^{T}$.  

$\tilde{H}_{1}$ is defined in direct analogy with $H_{1}$: just as $H_{1}$ is the subgroup of $F_{4}$ transformations (of $h_{3}(\mathbb{O})$) that preserve the embedding of $\mathbb{C}$ in $\mathbb{O}$, $\tilde{H}_{1}$ is the subgroup of $E_{6}$ transformations (of $h_{3}^{C}(\mathbb{O})$) that preserve the embedding of $\mathbb{C}$ in $\mathbb{O}$.   Let us describe this subgroup concretely.  As before, we can write an element $y\in h_{3}^{C}(\mathbb{O})$ as in Eq.~(\ref{Jordan_element}), except now the three $\alpha_{i}$ and $x_{i}$ are elements of $\mathbb{R}^{C}$ and $\mathbb{O}^{C}$ (the complexifications of $\mathbb{R}$ and $\mathbb{O}$); and, as before, we can split each such element $y$ into $\hat{y}+m$, except now $\hat{y}$ and $m$ are elements of $h_{3}^{C}(\mathbb{C})$ and $M_{3}^{C}(\mathbb{C})$ (the complexifications of $h_{3}(\mathbb{C})$ and $M_{3}(\mathbb{C})$).  Now, if we define
\begin{equation}
  \iota\equiv\frac{1}{2}(1+Il)\qquad{\rm and}\qquad
  V(V_{L},V_{R})\equiv V_{L} \bar{\iota}+V_{R}\iota
\end{equation}
then the subgroup $\tilde{H}_{1}$ of the $E_{6}$ transformations of $h_{3}^{C}(\mathbb{O})$ that preserve the embedding of $\mathbb{C}$ in $\mathbb{O}$ is given (Thrm. 3.13.5 in \cite{Yokota}) by the following generalization of Eq.~(\ref{H1}) :
\begin{equation}
  \label{Htilde1}
  \hat{h}\!\to\!V(V_{L},V_{R})\hat{h}V(V_{L},V_{R})^{\dagger},\quad m\!\to\!Um\bar{V}(V_{L},V_{R})^{\dagger}
\end{equation}
where $U$, $V_{L}$ and $V_{R}$ are arbitrary elements of $SU(3)$ ($3\times 3$ matrices whose elements are valued in the octonion sub-algebra $\mathbb{C}=\{a_{0}+a_{4}l\}$).  Since this transformation is unchanged by the transformation $\{U,V_{L},V_{R}\}\to {\rm e}^{2\pi i/3}\{U,V_{L},V_{R}\}$, these automorphisms form the subgroup $\tilde{H}_{1}=[SU(3)\times SU(3)\times SU(3)]/\mathbb{Z}_{3}$.

$\tilde{H}_{2}$ is defined in direct analogy with $H_{2}$: just as $H_{2}$ is the subgroup of $F_{4}$ transformations that preserve the rank-one idempotent (\ref{p1}), $\tilde{H}_{2}$ is the subgroup of $E_{6}$ transformations that preserve the rank-one idempotent (\ref{p1}).  This subgroup is $\tilde{H}_{2}=Spin(10)$ (Thrm 3.10.4 in \cite{Yokota}).

\subsection{LR-symmetric gauge group: $G_{LR}$.} 

In the previous section, we reviewed the main result in \cite{Todorov:2018mwd}: within $h_{3}(\mathbb{O})$, $H_{1}\cap H_{2}=G_{SM}$.  Now let us consider the analogous intersection $\tilde{H}_{1}\cap\tilde{H}_{2}$: these are the transformations of the form (\ref{Htilde1}) with the additional property that $A$ and $B$ both have the block diagonal form:
\begin{equation}
  V_{L}=\left(\begin{array}{c|c} \varphi\,v_{L} & 0 \\ \hline 0 & \varphi^{-2} \end{array}\right)\qquad
  V_{R}=\left(\begin{array}{c|c} \varphi\,v_{R} & 0 \\ \hline 0 & \varphi^{-2} \end{array}\right)
\end{equation}
where $v_{L},v_{R}\in SU(2)$ and $\varphi\in U(1)$.   Thus, an element $\{U,v_{L},v_{R},\varphi\}\in SU(3)\times SU(2)_{L}\times SU(2)_{R}\times U(1)$ determines a transformation in $\tilde{H}_{1}\cap\tilde{H}_{2}$; but this transformation is unchanged by the replacement $\{U,v_{L},v_{R},\varphi\}\to\{U,-v_{L},-v_{R},,-\varphi\}$ or the replacement $\{U,v_{L},v_{R},\varphi\}\to\{{\rm e}^{2\pi i/3}U,v_{L},v_{R},{\rm e}^{2\pi i/3}\varphi\}$.  So $\tilde{H}_{1}\cap\tilde{H}_{2}=G_{LR}$ where
\begin{equation}
  \label{G_LR}
  G_{LR}=[SU(3)\times SU(2)_{L}\times SU(2)_{R}\times U(1)]/\mathbb{Z}_{6}
\end{equation}
is the widely-studied minimal left-right-symmetric extension of $G_{SM}$.  This is one of our key new results.  
  
\subsection{LR-symmetric representation: $\rho_{LR}$.}  

Under $E_{6}$, the elements of $h_{3}^{C}(\mathbb{O})$ transform as a single complex irrep, the ${\bf 27}$ of $E_{6}$.  When restricted to the subgroup $\tilde{H}_{2}=Spin(10)$, ${\bf 27}$ splits into ${\bf 1}\oplus{\bf 10}\oplus{\bf 16}$ as follows:
\begin{equation}
  \label{Jordan_element}
  y=\left(\begin{array}{cc|c}
  \textcolor{red}{\alpha_{1}} & \textcolor{red}{x_{3}} & \textcolor{violet}{x_{2}^{\ast}} \\
  \textcolor{red}{x_{3}^{\ast}} & \textcolor{red}{\alpha_{2}} & \textcolor{violet}{x_{1}} \\
  \hline
  \textcolor{violet}{x_{2}} & \textcolor{violet}{x_{1}^{\ast}} & \textcolor{blue}{\alpha_{3}} \end{array}\right)
  \qquad(\alpha_{i}\in\mathbb{R}^{C},\;\;x_{i}\in\mathbb{O}^{C}).
\end{equation}
In particular, the column (or row) in violet transforms as the ${\bf 16}$ of $Spin(10)$ -- {\it i.e.} just like a single generation of fermions in $Spin(10)$ grand unification.  When we further restrict from $Spin(10)$ to its subgroup $\tilde{H}_{1}\cap\tilde{H}_{2}=G_{LR}$, the ${\bf 16}$ of $Spin(10)$ further splits into the representation
\begin{eqnarray}
  \label{rho_LR}
  \rho_{LR}\!&\!=\!&\!(3,2,1,+\frac{1}{6})\oplus(\bar{3},1,2,-\frac{1}{6}) \nonumber \\
  \!&\!\oplus\!&\!(1,2,1,-\frac{1}{2})\oplus(1,1,2,+\frac{1}{2}).
\end{eqnarray}

\subsection{LR-symmetric model.}  

But $G_{LR}$ (\ref{G_LR}) and $\rho_{LR}$ (\ref{rho_LR}) are precisely the gauge group and fermion representation in the minimal left-right symmetric extension of the standard model \cite{Mohapatra:1974hk}.  A particularly appealing version of this model \cite{Hall:2018let, Hall:2019qwx, Dror:2020jzy} has two Higgs fields, $H_{L}$ and $H_{R}$, matter content neatly summarized as follows
\begin{equation}
  \label{SM_table}
  \begin{array}{c|c|c|c|c} 
    & SU(3) & SU(2)_{L} & SU(2)_{R} & U(1) \\
    \hline
    q_{L}^{i} & 3 & 2 & 1 & +1/6 \\
    \hline
    q_{R}^{i} & 3 & 1 & 2 & +1/6 \\
    \hline
    l_{L}^{i}, H_{L} & 1 & 2 & 1 & -1/2 \\
    \hline
    l_{R}^{i}, H_{R} & 1 & 1 & 2 & -1/2 
  \end{array}
\end{equation}
and a $\mathbb{Z}_{2}$ symmetry under the combined action of spatial parity and the exchange $\{q_{L}^{i},l_{L}^{i},H_{L}\}\leftrightarrow\{q_{R}^{i},l_{R}^{i},H_{R}\}$.  If the VEV of $H_{R}$ is much higher than the electroweak scale, then below this scale, the theory reduces to the standard model, with gauge group $G_{SM}$, fermion representation $\rho_{SM}$, and the usual Higgs field ($H_{L}$).  As shown in \cite{Hall:2018let, Hall:2019qwx, Dror:2020jzy}, this model is not only experimentally viable, but can simultaneously:  (i) explain the vanishing of the Higgs coupling $\lambda$ at $\sim10^{10}~{\rm GeV}$; (ii) provide an elegant solution to the strong-CP problem \cite{Babu:1989rb}; (iii) give precise gauge-coupling unification; and (iv) account for dark matter and the cosmological matter/anti-matter asymmetry.

\section{The magic square and triality.}  The preceding ideas may be given a geometric interpretation, via the intriguing ``magic square" construction \cite{Baez:2001dm}.  

The magic square maps each pair of normed division algebras $(\mathbb{K}, \tilde{\mathbb{K}})$ to a Lie group $M(\mathbb{K},\tilde{\mathbb{K}})$, which may be interpreted as the symmetry group of the ``Rosenfeld projective plane" $(\mathbb{K}\otimes\tilde{\mathbb{K}})\mathbb{P}^{2}$ \cite{Baez:2001dm}.   The construction may be elegantly formulated in terms of triality: the Lie algebra $\mathfrak{m}(\mathbb{K},\tilde{\mathbb{K}})$ corresponding to $M(\mathbb{K},\tilde{\mathbb{K}})$ has the form \cite{Ramond, BartonSudbery}
\begin{equation}
  \label{triality_0}
  \mathfrak{m}(\mathbb{K},\tilde{\mathbb{K}})\!=\!\mathfrak{t}(\mathbb{K})\!\;\!+\!\;\!\mathfrak{t}(\tilde{\mathbb{K}})\!\;\!+\!\;\!(\mathbb{K}\otimes\tilde{\mathbb{K}})_{\!\;\!1}\!+\!\;\!(\mathbb{K}\otimes\tilde{\mathbb{K}})_{\!\;\!2\;\!}\!+\!\;\!(\mathbb{K}\otimes\tilde{\mathbb{K}})_{\!\;\!3},\!\!
\end{equation}
where $\mathfrak{t}(\mathbb{K})$ is the triality algebra of $\mathbb{K}$, and the Lie bracket on the right-hand side is defined {\it e.g.}\ in Section 4.3 of \cite{BartonSudbery}.

In particular, $E_{6}$ arises as $M(\mathbb{C},\mathbb{O})$ and is interpreted as the symmetry group of the complex octonionic projective plane $(\mathbb{C}\otimes\mathbb{O})\mathbb{P}^{2}$.  Using $\mathfrak{t}(\mathbb{C})\!=\!\mathfrak{u}(1)+\mathfrak{u}(1)$ and $\mathfrak{t}(\mathbb{O})\!=\!\mathfrak{so}(8)$, Eq.~(\ref{triality_0}) then becomes \cite{Baez:2001dm, Adams}
\begin{subequations}
  \begin{eqnarray}
    \label{triality_1}
    \mathfrak{e}_{6}\!\!&\!=\!&\!\!\textcolor{blue}{\mathfrak{u}(1)}\!+\!\textcolor{red}{\mathfrak{so}(2)}\!+\!\textcolor{red}{\mathfrak{so}(8)}\!+\!\textcolor{red}{(\mathbb{C}\!\otimes\!\mathbb{O})}
    \!+\!\textcolor{violet}{(\mathbb{C}\!\otimes\!\mathbb{O})^{2}}\qquad \\
    \label{triality_2}
    \!\!&\!=\!&\!\!\textcolor{blue}{\mathfrak{u}(1)}+\textcolor{red}{\mathfrak{so}(10)}+\textcolor{violet}{(\mathbb{C}\otimes\mathbb{O})^{2}}.
  \end{eqnarray}
\end{subequations}

To interpret this equation, we recall the perspective of Klein (or Cartan) geometry \cite{Sharpe, Wise:2006sm}, where a symmetric space is a coset space $X=G/H$, with $G$ the symmetry group of $X$, and $H$ the subgroup that stabilizes a point.

First recall how this perspective applies to (``external") spacetime geometry $M_{4}$: we can regard Minkowski space $M_{4}$ as $G_{ext}/H_{ext}$, where $G_{ext}=ISO(3,1)$ is the Poincare group, and $H_{ext}=SO(3,1)$ is the Lorentz group. At the Lie algebra level, we start from the full Poincare algebra $\mathfrak{iso}(3,1)=\mathfrak{so}(3,1)+\mathfrak{p}$, remove the stabilizer subalgebra $\mathfrak{so}(3,1)$, and then the remaining translation generators $\mathfrak{p}$ form the tangent space to $M_{4}$, and automatically transform as the 4d (vector) irrep of $H_{ext}=SO(3,1)$.

The same perspective provides a closely parallel account of the (``internal") geometry $(\mathbb{C}\otimes\mathbb{O})\mathbb{P}^{2}$: we can regard the complex octonionic projective plane $(\mathbb{C}\otimes\mathbb{O})\mathbb{P}^{2}$ as $G_{int}/H_{int}$, where $G_{int}=E_{6}$, and $H_{int}=[U(1)\times Spin(10)]/\mathbb{Z}_{4}$ \cite{Baez:2001dm, Yokota}.  At the Lie algebra level, we start from Eq.~(\ref{triality_2}) for $\mathfrak{e}_{6}$, remove the stabilizer subalgebra $\mathfrak{u}(1)+\mathfrak{so}(10)$, and then the remaining generators $(\mathbb{C}\otimes\mathbb{O})^{2}$ form the tangent space to $(\mathbb{C}\otimes\mathbb{O})\mathbb{P}^{2}$, and automatically transform as the 16d irrep of $H_{int}$.  

Note that $H_{int}$ is the subgroup of $G_{int}$ that stabilizes the rank-one idempotent $\Pi$ in Eq.~(\ref{p1}) up to a phase (see Lemma 3.10.1 in \cite{Yokota}) while, as before, $Spin(10)< H_{int}$ is the subgroup that stabilizes $\Pi$ precisely, and $G_{LR}<Spin(10)$ is the subgroup that {\it also} stabilizes the embedding of $\mathbb{C}$ in $\mathbb{O}$.  Thus, the tangent space ($\mathbb{C}\otimes\mathbb{O})^{2}$ of $(\mathbb{C}\otimes\mathbb{O})\mathbb{P}^{2}$ transforms precisely as the ${\bf 16}$ of $Spin(10)$, as the representation $\rho_{LR}$ of $G_{LR}$, and as the representation $\rho_{SM}$ of $G_{SM}$ -- {\it i.e.}\ just as desired for a single generation of standard model fermions.

Much as $H_{int}=SO(3,1)$ is promoted to a local lorentz symmetry, mixing the four components in a tetrad, $H_{ext}$ (or $Spin(10)$, or $G_{LR}$, or $G_{SM}$) is promoted to a local gauge symmetry, mixing the 16 fermions in a generation. 

Note that, in rewriting (\ref{triality_0}) in the form (\ref{triality_2}) we had to choose one of the three copies of $(\mathbb{C}\otimes\mathbb{O})$ in (\ref{triality_1}) to include in $\mathfrak{so}(10)$; but these three copies of $(\mathbb{C}\otimes\mathbb{O})$ are permuted by $\mathfrak{so}(8)$ triality symmetry in (\ref{triality_1}) \cite{Baez:2001dm, BartonSudbery, Ramond}.  In this sense, if the standard model fermions really correspond to the tangent space $(\mathbb{C}\!\otimes\!\mathbb{O})^{2}$ of $(\mathbb{C}\otimes\mathbb{O})\mathbb{P}^{2}$, as we have suggested, then they inevitably arise in three triality-related ways when constructing $(\mathbb{C}\otimes\mathbb{O})\mathbb{P}^{2}$ from $E_{6}$.  It is natural to suspect that this is the origin of the three generations of fermions in the standard model.
%\footnote{This topic will be further explored in a follow-up paper.}

Many questions remain about the ideas presented here, but we feel our results already provide intriguing evidence that $h_{3}^{C}(\mathbb{O})$, $(\mathbb{C}\otimes\mathbb{O})\mathbb{P}^{2}$, and the related objects discussed above, are intimately connected to the standard model.  

{\it Note added:} Let us mention two interesting related papers which provide an important complement to the perspective given here.  The first paper \cite{Krasnov:2019auj} shows how, rather than starting from the $3\times 3$ octonionic model of $F_{4}$ based on the exceptional Jordan algebra $h_{3}(\mathbb{O})$, then obtaining $Spin(9)$ as the subgroup of $F_{4}$ which fixes a copy of $h_{2}(\mathbb{O})$ within $h_{3}(\mathbb{O})$, and finally obtaining $G_{SM}$ as the subgroup of this $Spin(9)$ that further fixes an embedding of $\mathbb{C}$ in $\mathbb{O}$ (as done in Ref.~\cite{Dubois-Violette:2018wgs} and reviewed in Sec.~II above), one can instead start directly from the $2\times 2$ octonionic model of $Spin(9)$ based on $h_{2}(\mathbb{O})$ and then obtain $G_{SM}$ as the subgroup of $Spin(9)$ that commutes with a certain complex structure in the space of $Spin(9)$ spinors.  Similarly, the second paper \cite{Krasnov:2022meo} (which appeared after the present paper appeared on the arXiv) shows how, rather than starting from the $3\times3$ octonionic model of $E_{6}$ based on the complexified exceptional Jordan algebra $h_{3}^{C}(\mathbb{O})$, then obtaining $Spin(10)$ as the subgroup that fixes a copy of $h_{2}^{C}(\mathbb{O})$ within $h_{3}^{C}(\mathbb{O})$, and finally obtaining $G_{LR}$ as the subgroup of this $Spin(10)$ that further fixes an embedding of $\mathbb{C}$ in $\mathbb{O}$ (as done in Sec.~III above), one can instead start directly from the $2\times2$ octonionic model of $Spin(10)$ based on $h_{2}^{C}(\mathbb{O})$ and then obtain $G_{LR}$ as the subgroup of $Spin(10)$ that commutes with a certain complex structure on the space of $Spin(10)$ semi-spinors (and, moreover, obtain $G_{SM}$ as the subgroup of $Spin(10)$ that commutes with a certain appropriately-aligned commuting pair of such complex structures).  In this sense, the story (at least for a single generation) may be phrased purely in terms of $2\times2$ octonionic structures, without mentioning the $3\times3$ exceptional Jordan algebras.

\begin{acknowledgments}
I thank Ben Albert, Shane Farnsworth and Ivan Todorov for valuable conversations.  This paper is dedicated to the memory of Ivan Todorov.  LB is supported by the STFC Consolidated Grant ``Particle Physics at the Higgs Centre."  Research at PI is supported by the Government of Canada through the Department of Innovation, Science \& Economic Development and the Province of Ontario via the Ministry of Research \& Innovation.
\end{acknowledgments}

\end{document}